\documentclass[12pt,onecolumn,draftcls]{IEEEtran}
\usepackage{cite}
\ifCLASSINFOpdf
   \usepackage[pdftex]{graphicx}
  \graphicspath{{../pdf/}{../jpeg/}}
  \DeclareGraphicsExtensions{.pdf,.jpeg,.png}
\else
   \usepackage[dvips]{graphicx}
   \graphicspath{{../eps/}}
  \DeclareGraphicsExtensions{.eps}
\fi
\usepackage[cmex10]{amsmath}
\usepackage{amsfonts}
\usepackage{amssymb}
\usepackage{mathrsfs}
\usepackage{array}
\usepackage{mdwmath}
\usepackage{mdwtab}
\usepackage{eqparbox}
\usepackage[tight,footnotesize]{subfigure}
\usepackage{fixltx2e}
\usepackage{stfloats}
\usepackage{url}
\newtheorem{theorem}{Theorem}[section]

\newcommand{\qed}{\nobreak \ifvmode \relax \else
      \ifdim\lastskip<1.5em \hskip-\lastskip
      \hskip1.5em plus0em minus0.5em \fi \nobreak
      \vrule height0.75em width0.5em depth0.25em\fi}

\makeatletter

\newcommand{\Rmnum}[1]{\expandafter\@slowromancap\romannumeral #1@}
\makeatother

\begin{document}
\title{Sparse Channel Estimation by Factor Graphs}
\author{Rad~Niazadeh,~\IEEEmembership{Member,~IEEE,}~Massoud~Babaie-Zadeh,~\IEEEmembership{Senior Member,~IEEE,} and~\\Christian~Jutten,~\IEEEmembership{Fellow,~IEEE}
\thanks{* This work has been partially funded by Iran Telecom Research Center (ITRC), and also by center for International Research and Collaboration (ISMO) and French embassy in Tehran in the framework of a GundiShapour collaboration program.}}

\maketitle
\begin{abstract}
The problem of estimating a sparse channel, i.e. a channel
with a few non-zero taps, appears in various areas of communications. Recently, we
have developed an algorithm based on iterative alternating minimization which iteratively detects the location and the value of the
taps. This algorithms involves an approximate Maximum
A Posteriori (MAP) probability scheme for detection of the location of taps, while
a least square method is used for estimating the values at
each iteration. In this work, based on the method of factor graphs and message passing algorithms, we will compute an exact solution for the MAP estimation problem. Indeed, we first find a factor graph model of this problem, and then perform the well-known min-sum algorithm on the edges of this graph. Consequently, we will find an exact estimator for the MAP problem that its complexity grows linearly with respect to the channel memory. By substituting this estimator in the mentioned alternating minimization method, we will propose an estimator that will nearly achieve the Cram\'{e}r-Rao bound of the genie-aided estimation of sparse channels (estimation based on knowing the location of non-zero taps of the channel), while it can perform faster than most of the proposed algorithms in literature. 
\end{abstract}
\section{Introduction}
The problem of sparse channel estimation, i.e. estimation of a channel whose impulse response has a few non-zero taps, has a broad scope of applications in many areas of communication such as UWB communications and acoustic underwater transmissions~\cite{carbonelli2007sparse,kocic1995sparse}. Indeed, this problem has an even earlier origin in geology, where the problem of sparse channel estimation was introduced for finding the caste structure of earth in an specific area using seismic signals~\cite{chapman1983deconvolution}. In this problem, it is assumed that we have a sparse channel for communication between a transmitter and receiver. At the first step, we send a training sequence from the transmitter, which is known at the receiver side, and then we estimate the value and locations of non-zero taps of the mentioned sparse channel based on observing the output of the channel. Mathematically, we have the following model:
\begin{equation}
\label{eq1}
y(k)=u(k)*h(k)+n(k)\enspace ,
\end{equation}
in which $y(k)$ is the output signal of the channel, $h(k)$ is the $K$-sparse\footnote{By $K$-sparse we mean that there are at most $K$ non-zero elements in $\mathbf{h}=[h(0),h(1),\dots,h(M-1)]^T$.} channel with $M$ taps, $u(k)$ is the $L$-length training sequence at the input of the channel and $n(k)$ is a white Gaussian noise. By observing all of the $N=M+L-1$ output symbols of the channel and by having the assumption of $L\leq M$, we have the following model in matrix form:
\begin{equation}
\label{eq2}
\mathbf{y}=\mathbf{U}\mathbf{h}+\mathbf{n}=\mathbf{U_h}\mathbf{b}+\mathbf{n}\enspace ,
\end{equation}
in which $\mathbf{y}\triangleq[y(0),y(1),\dots,y(N-1)]^T$ is the $N \times 1$ vector of output symbols of the channel, $\mathbf{n}\triangleq[n(0),n(1),\dots n(N-1)]^T$ is an $N\times 1$ white Gaussian noise vector (i.e. $\mathbf{n}\sim N(\sigma^2\mathbf{I}_N,0)$), and $\mathbf{U}$ is an $N\times M$ full-rank training matrix, defined as follows:
\begin{equation}
\label{eq3}
\mathbf{U}_{L+M-1\times M}\triangleq[\mathbf{\mathscr{U}}^{(0)},\mathbf{\mathscr{U}}^{(1)},\dots,\mathbf{\mathscr{U}}^{(M-1)}]\enspace ,
\end{equation}
in which, $\mathbf{\mathscr{U}}^{(i)}$ is the $i$-th shift of the vector $\mathbf{\mathscr{U}}^{(0)}_{(L+M-1)\times 1}=[u(0),u(1),\dots,u(L-1),0,\dots,0]^T$. Additionally, $\mathbf{U_h}\triangleq\mathbf{U}\mbox{diag}(\mathbf{h})$ and $\mathbf{b}\in\{0,1\}^{M}$ is a binary vector that indicates the locations of non-zero taps, i.e.
\begin{equation}
\label{eq4}
\forall i\in\{1,2,\dots M\}:~~~b_i=\left\{
\begin{array}{rl}
1 & \text{if } h_i\neq 0,\\
0 & \text{if }h_i=0.
\end{array} \right.
\end{equation}   
As it can be seen from (\ref{eq2}), sparse channel estimation is indeed an overdetermined and noisy sparse recovery problem. The goal is to find an estimate $\mathbf{\hat{h}}$ based on the observation vector $\mathbf{y}$ with a Mean Square Error (MSE), which is defined as the value of $\mathbb{E}\{\lVert\mathbf{h}-\mathbf{\hat{h}}\rVert^2\}$, as low as possible. On the one hand, it is well known that the Least Square Estimator (LSE), which is defined as follows:
\begin{equation}
\label{eq5}
\mathbf{\hat{h}}=\underset{\mathbf{h}}{\mbox{argmin}}\lVert\mathbf{y}-\mathbf{U}\mathbf{h}\rVert_2^2\enspace ,
\end{equation}
 is the optimal estimator in the sense of MSE when the estimator does not have any prior knowledge about neither the sparsity structure of $\mathbf{h}$ (i.e. the location of non-zero taps), nor its degree of sparsity (i.e. $K$)~\cite{sharp2008estimation,carbonelli2007sparse}. Moreover, LSE will achieve the Cram\'{e}r-Rao lower bound of the estimation of $\mathbf{h}$ based on $\mathbf{y}$ without any prior information around $\mathbf{h}$~\cite{cotter2002sparse,sharp2008estimation,carbonelli2007sparse}. In literature, this bound, which is indeed the Cram\'{e}r-Rao lower bound of the unstructured estimation problem, is known as CRB-US\cite{carbonelli2007sparse,niazadeh2010alternating}. On the other hand, if a genie aids us with the location of the non-zero taps of $\mathbf{h}$, then it can be shown that~\cite{niazadeh2010alternating,carbonelli2007sparse} the Structured Least Square Estimator (SLSE), which finds the solution of the following problem:
\begin{equation}
\label{eq6}
\mathbf{\hat{h}_\tau}=\underset{\mathbf{h_\tau}}{\mbox{argmin}}\lVert\mathbf{y}-\mathbf{U_\tau}\mathbf{h_\tau}\rVert_2^2\enspace ,
\end{equation}
will be an efficient estimator~\cite{papoulis2002probability,kay1998fundamentals}. In (\ref{eq6}), $\tau\subset\{0,1,\dots M-1\}$ is the support of $\mathbf{h}$, $\mathbf{U_\tau}$ is a sub-matrix of $\mathbf{U}$ that includes columns corresponding to the indices in $\tau$ and $\mathbf{h_\tau}$ is a $K\times 1$ vector that contains the non-zero taps of $\mathbf{h}$. Here, by \textit{efficient} we mean that this estimator will achieve Cram\'{e}r-Rao lower bound of the channel estimation problem, in which the receiver estimates $\mathbf{h}$ based on $\mathbf{y}$ and $\tau$. In literature, this bound, which is indeed the Cram\'{e}r-Rao lower bound of the genie aided (structured) estimation problem, is known as CRB-S\cite{carbonelli2007sparse,babadi2009asymptotic}.

In literature, the problem of finding an estimator that achieves the CRB-S without a complete knowledge about the structure of the sparse channel has been in the point of interest. The goal is to design a practical estimator with MSE as close as possible to that of SLSE, while this estimator has no prior knowledge about the sparsity structure of the channel and it just knows the degree of sparsity, i.e. $K$. Cand\`es \textit{et al.} \cite{candes2007dantzig} and Haupt \textit{et al.}~\cite{haupt2006signal} proposed estimators that can achieve CRB-S to a factor of $\log M$. Similarly, Babadi \textit{et al.}~\cite{babadi2009asymptotic} showed that with the use of an estimator known as \textit{Typical} Estimator we can asymptotically achieve the Cram\'{e}r-Rao bound of the genie aided estimation. Additionally, the work done by Carbonelli \textit{et al.}~\cite{carbonelli2007sparse} proposed practical algorithms for this problem which can reach bounds close to CRB-S in the MSE sense. In our previous work, we have also proposed a family of estimators for this problem based on the method of alternating minimization~\cite{niazadeh2010alternating}, in which we jointly estimate the location and value of non-zero taps of the channel.

In~\cite{niazadeh2010alternating}, very similar to the idea in~\cite{carbonelli2007sparse} and~\cite{cotter2002sparse}, we have used an iterative procedure based on alternating minimization for jointly finding both $\mathbf{h}$ and $\mathbf{b}$. At each iteration, we first assume that we have an appropriate estimate for $\mathbf{b}$ i.e. $\mathbf{\hat{b}}$, then we estimate $\mathbf{h}$ at this iteration by finding the solution of the following problem:
\begin{equation}
\begin{split}
\label{eq7}
\mathbf{\hat{h}}=\underset{\mathbf{h}}{\mbox{argmin}}~\lVert \mathbf{y}-\mathbf{U}\mbox{diag}(\mathbf{\hat{b}})\mathbf{h}\rVert_2=\underset{\mathbf{h}}{\mbox{argmin}}~\lVert \mathbf{y}-\mathbf{U_{\hat{b}}}\mathbf{h}\rVert_2=\mathbf{U_{\hat{b}}}^\dag\mathbf{y} \enspace,
\end{split}
\end{equation}
in which $(.)^\dag$ denotes the Moore-Penrose pseudo-inverse operator. Then we will try to find a more accurate estimate for $\mathbf{b}$ at this iteration based on $\mathbf{\hat{h}}$, and hence we perform a MAP estimation for $\mathbf{b}$ based on the observation $\mathbf{y}\approx\mathbf{U_{\hat{h}}}\mathbf{b}+\mathbf{n}$.
For the MAP estimation, we need prior knowledge on the probability distribution of $\mathbf{b}$. By assuming the simple model of i.i.d Bernoulli distribution for the elements of channel location vector, i.e.\{${b_i}\}_{i=1}^M$, and by defining $P_a=\frac{K}{M}<\frac{1}{2}$, then we can assume that $\mathbb{P}\{b_i=1\}=P_a$ and so:
\begin{equation}
\label{eq7}
\mathbb{P}\{\mathbf{b}\}=\prod_{i=1}^{M}{\mathbb{P}\{b_i\}}=(1-P_a)^{(M-\lVert\mathbf{b}\rVert_0)}P_a^{\lVert \mathbf{b}\rVert_0}\enspace .
\end{equation}
Now, as was mentioned in~\cite{niazadeh2010alternating}, we can find the MAP solution as:
\begin{align}
\mathbf{\hat{b}}_{\textrm{MAP}}&=\underset{\mathbf{b}\in\{0,1\}^M }{\mbox{argmax}}~\mathbb{P}\{\mathbf{y}|\mathbf{\hat{h}},\mathbf{b}\}\mathbb{P}\{\mathbf{b}\}
=\underset{\mathbf{b}\in\{0,1\}^M }{\mbox{argmax}}~\exp{(-\frac{1}{2\sigma^2}\lVert\mathbf{y}-\mathbf{U_{\hat{h}}}\mathbf{b}\rVert_2^2)}(\frac{P_a}{1-P_a})^{\lVert \mathbf{b}\rVert_0}\nonumber\\
&\label{eq9}=\underset{\mathbf{b}\in\{0,1\}^M }{\mbox{argmin}}~\lVert\mathbf{y}-\mathbf{U_{\hat{h}}}\mathbf{b}\rVert_2^2+\lambda \lVert \mathbf{b}\rVert_0 \enspace,
\end{align}
in which $\lambda=2\sigma^2\ln(\frac{1-P_a}{P_a})>0$. It is important to mention that the direct solution of (\ref{eq9}) requires a combinatorial search over $\{0,1\}^M$ for finding $\mathbf{\hat{b}}_{\textrm{MAP}}$, and hence it complexity grows exponentially with respect to $M$. Accordingly, in~\cite{niazadeh2010alternating}, we had focused on finding an approximate solution to (\ref{eq9}) instead of the direct search, which would have much less complexity order in comparison to this mentioned direct search.

In this paper, we concentrate on the problem of finding the \textit{exact} solution of (\ref{eq9}) based on the method of factor graphs and the associated summary propagation algorithms~\cite{kschischang2001factor,loeliger2004introduction,colavolpe2005application}. Factor Graphs (FG) are bipartite graphs that can be used for modelling different problems that arise in various areas such as coding, detection theory, and social sciences~\cite{kschischang2001factor}. Moreover, by the use of summary propagation algorithms~\cite{kschischang2001factor} along the edge of the graph, one may find an algorithm that can perform in polynomial time with respect to some model parameters and can find the exact solution of a discrete optimization problem that may be solved directly by an exponentially order algorithm. An example is the problem described in equation (\ref{eq9}), and further examples can be find in coding, where Viterbi algorithm is used for decoding of convolutional codes~\cite{lin1983error}~\footnote{It is also important to note that the problem in (\ref{eq9}) is \textit{not} a NP hard problem, as the problem of decoding of convolutional codes is not, and just the method of direct search could be performed in a non-polynomial time. Indeed, the role of our algorithm in finding the \textit{exact} solution of (\ref{eq9}) is similar to the role of Viterbi algorithm in finding the \textit{exact} solution of maximum likelihood decoding of convolutional coeds.}. Two main summary propagation algorithms are the sum-product (or belief propagation or probability propagation) and the min-sum (or max-Sum) algorithms, both of which have a lot of applications in the context of error correcting codes, finance, and dynamic programming~\cite{loeliger2004introduction}. In this work, we first find the factor graph model of the discrete optimization problem described in (\ref{eq9}), and then will try to perform the min-sum algorithm on the edges of this graph in order to find a MAP estimator that can perform in polynomially order with respect to channel memory, i.e. $M$. Afterwards, we will try to use this exact MAP estimator in our alternating minimization algorithm for estimating sparse channels, and will obtain a near-optimal practical algorithm that can nearly achieve CRB-S. 

This paper is organized as follows. In Section~\ref{sec2}, we will have a summary of important notions and concepts from the context of factor graphs and summary propagation algorithms, which operate by passing ``messages" along the edges and ``summarizing" them at each node. Afterwards, in Section~\ref{sec3},  we will introduce the factor graph model of the problem of MAP estimation of the location of non-zero taps. In this section, we will also propose a method based on the min-sum algorithm that operates on the factor graph model of Section~\ref{sec2} and finds the exact solution of (\ref{eq9}). Then we will introduce a new variant of our alternating minimization algorithm in~\cite{niazadeh2010alternating} that uses this mentioned method for finding the MAP solution. Finally, we will study the behaviour of our proposed algorithm in the sense of MSE using simulation. In this part, we will see that our algorithm will reach nearly to CRB-S, while the computation cost of its implementation is increasing polynomially with respect to channel memory. Moreover, we will see that the computational complexity of our estimator will increase exponentially with respect to the length of the training sequence.

\section{\label{sec2}A Review of Factor Graphs and Message-Passing Algorithms}
In this section, we will try to have a brief review on factor graphs and some associated concepts for the sake of readability, which is mainly based on the works of Kschischang \textit{et al.}~\cite{kschischang2001factor}. At first, we review the concept of factor graphs and some useful notations and definitions, such as cross sections and summary operation. Then the operation of sum-product algorithm (as an instance of message-passing algorithms) on the edges of a factor graph, as well as its variants and proposed scheduling methods for performing this algorithm will be discussed.
\subsection{Notations}
A factor graph is a bipartite graph that describes how a ``global" function of multiple variable factorizes into a product of ``local" functions~\cite{kschischang2001factor}. More precisely, assume that $X_S=\{x_i:i\in S\}$ is a set of variables whose indices belong to a finite set of indices $S=\{i_1,i_2,\dots i_N\}$, in which $i_1\leq i_2\leq \dots \leq i_N$. Moreover, assume that for every $i\in S$, $x_i$ takes its value from a set $A_i$. We also define $X_E=\{x_i:i\in E\subset S\}$ for every $E\subset S$. A particular assignment of a value to each of the variables of $X_S$ such as $a=(a_{i_1},a_{i_2},\dots,a_{i_N})$ will be called a \textit{configuration} of variables, and this assignment ($x_{i_j}=a_{i_j}$ for every $1\leq j \leq N$) is denoted by $X_S=a$. The set of all possible configurations is equal to the Cartesian product $A_S=\prod_{i\in S}{A_i}$. Obviously, a configuration is an element of $A_S$. In a similar way, for a subset of indices like $E=\{j_1,j_2,\dots j_M \}\subset S$ (in which $j_1\leq j_2\leq \dots \leq j_M$) and a configuration of variables like $a$ in $S$, we can define the sub-configuration of $a$ with respect to $E$ as $a_E=(a_{j_1},a_{j_2},\dots a_{j_M})$. Each of these sub-configurations is an element of the Cartesian Product $A_E=\prod_{i\in E}{A_i}$ . Now, assume that we have $g:A_S\rightarrow R$ as a function of variables in $X_S$ and with a codomain equal to an arbitrary discrete or continuous set $R$. At this state, we require that a binary product (denoted by `.') and a unit element (denoted by $1$) is defined in $R$ that satisfy the following equation for every $u$,$v$ and $w$ in $R$,
\begin{equation}
\label{eq10}
1.u=u~~~,~~~u.v=v.u~~~,~~~(u.v).w=u.(v.w)\enspace,
\end{equation}
so that $R$ is commutative semi-group. Further, assume that for a collection $Q$ of subsets of $S$, $g(X_S)$ factorizes as $g(X_S)=\prod_{E\in Q}f_E(X_E)$ in which for every $E\in Q$, $f_E: A_E\rightarrow R$ is a local function of variables in $X_E$. Now, the factor graph representation of this factorization is a bipartite graph denotes as $F(Q,S)$, in which $S\cup Q$ is the set of nodes and $\left \{\{i,E\}:i\in S,E\in Q, i\in E  \right\}$ is the set of edges. To simplify, we have a node for each variable and another node for each local function. Moreover, there exists an edge such as $\{i,E\}$ between a variable node, $x_i$, and a function node, $f_E$,  if $f_E(X_E)$ is a function of $x_i$, or equally $i\in E$. 

In the context of factor graphs, in addition to the ordinary terminology and ideas from graph theory (such as the definitions for node, edge, loop, path, tree, leaves of a tree and neighbours of a node~\cite{bondy1976graph}), there are some complementary notations. In this context, the variable node that is associated with the edge $\{u,v\}$ is also referred to as $x_{u,v}$. Moreover, we also need the notation of ``Iverson`s conventions"~\cite{iverson1964method} for indicating the truth of a logical proposition. In fact, if $P$ is a logical proposition, then $[P]$ is a binary indicator that shows whether if $P$ is true or not. So, similar to~\cite{kschischang2001factor}, we have:
\begin{equation}
\label{eq11}
[P] = \left\{
\begin{array}{rl}
1& ~~~~~ \text{if $P$, } \\
0& ~~ \text{if not $P$. }
\end{array} \right.
\end{equation}    
We will use these notations and definitions throughout the rest of this paper.

\subsection{Cross Sections and Their Associated Factor Graph}
In this section, we will review the concept of cross sections~\cite{kschischang2001factor} in a factor graph, which will play an important role in the later parts of this paper. Assume that $E\subset S$ is a subset of indices in $S$. Further, suppose that $X_E$ is a set of variables with indices in $E$ and $a_E$ is a specific configuration for $X_E$. Then, the cross section of $g(X_S)$ with respect to $X_E=a_E$, which is denoted by $g(X_{S\backslash E}||X_E=a_E)$ in~\cite{kschischang2001factor}, can be defined as following:
\begin{equation}
\label{eq12}
g(X_{S\backslash E}||X_E=a_E)=g(X_{S\backslash E})|_{X_E=a_E}~~.
\end{equation}
It is almost obvious that the corresponding factor graph representation of this cross section (factor graph of $g(X_{S\backslash E}||X_E=a_E)$) can be obtained in two steps. At the first step, we should delete all variable nodes of the factor graph representation of $g(X_S)$ that are associated with the indices in $E$ (all of nodes that represents the variables in $X_E$). Second, we should replace some of the function nodes, whose associated local function is a function of a variable in $X_E$, by their corresponding cross section, which can be obtained by assigning the value of variables in $X_E$ by $a_E$. 

It is important to mention that in many cases, such as ours in this paper, we deal with the factor graph representation of a probability density function of some random variables. In this case, if $g(X_S)$ is the probability density function of random variables in $X_S$, then the cross section $g(X_{S\backslash E}||X_E=a_E)$ is proportional to the conditional probability density function of variables in $X_{S\backslash E}$ under the condition of $X_E=a_E$~\cite{kschischang2001factor}. 

\subsection{Summary Operation and Marginal Functions}
In this section, we will review the definition of an important operator, named as summary operator~\cite{kschischang2001factor,colavolpe2005application,loeliger2004introduction}, 
which will play an important role in introducing the well-known sum-product algorithm.

 Assume that for every $i\in S$, $A_i$ is an alphabet with finite number of elements. By the use of an addition operator (which we denote by ``$+$"), we can define the summary operator with respect to a subset of variables in $X_S$. More accurately, we may define an addition operator that should be commutative and associative in the domain of $g(X_S)$ (the semi-group $R$)~\cite{kschischang2001factor}. Additionally, it is necessary that the product operator of this semi-group (which is denoted by ``$.$") satisfy the distributive property with respect to this addition operator. In mathematical words, for every $x,y,z\in R$ we should have~\cite{kschischang2001factor}:
\begin{equation} 
\label{eq13}
 x (y+z)=(x.y)+(x.z)~~.
\end{equation}
In this case, $(R,+,.)$ is a semi-ring~\cite{matsumura1989commutative,kschischang2001factor}. Now, for arbitrary subsets of $S$ like $B$ and $C$, the summary operator of variables in $X_B$ with respect to $X_C$, which is denoted by  $\downarrow X_C$, will be defined as following: 
\begin{equation}
\label{eq14}
g(X_B)\downarrow X_C=\sum_{x_i\in A_i :i\in B\cap (S\backslash C)}{g(X_B)}~~,
\end{equation}
in which $g(X_B)\downarrow X_C$ is the marginal function of $g(X_B)$  with respect to the variables in $X_{B\cap (S\backslash C)}$~\cite{kschischang2001factor}. As it can be seen, this marginal function is obtained by taking a sum of $g(X_B)$ over all possible configurations of variables that are in $X_B$, but are not in $X_C$. 

It is also important to note that if $g(X_S)$ is a probability density function (which is a function in $R=\mathbb{R}$), and add and product operators are ordinary add and product operator in $\mathbb{R}$, then $g(X_S)\downarrow x_i$ will be proportional to the marginal probability density function of variable $x_i$~\cite{kschischang2001factor}. Further, we can also consider another special case in which $R=\mathbb{R}$, ``$\textrm{min}$" operator is used in place of ``$+$", and the ordinary sum operator in $\mathbb{R}$ is used as ``$.$". Note that for every $x,y,z\in \mathbb{R}$ we have:
\begin{equation}
\label{eq15}
x+\textrm{{min}}(y,z)=\textrm{{min}}(x+y,x+z)~~,
\end{equation}  
and so the ordinary sum operator is associated in the ``$\textrm{min}$" operator. Thus, $(\mathbb{R},\textrm{min},+)$ is a semi-ring and can be used for defining a summary operator. In this case, $g(X_S)\downarrow X_B$ will be equal to the minimum of $g(X_S)$ over all possible configurations of $X_B$. This summary operator can be used for finding the solution of a discrete optimization problem, as will be used later in this paper.  
\subsection{Sum-Product Algorithm and Message Scheduling}
In this section, we will review the concept of message passing algorithms, and in particular sum-product algorithm, which is provided by Aji \textit{et al.} for the first time in 1997~\cite{aji2000generalized}. The foundation of this algorithm is based on exchange of so called ``messages" between the nodes of a factor graph, which will result in the computation of a marginal function. More accurately, in the progress of sum-product algorithm one can assume that we have a processor  at each node. In this model, every edge in the factor graph is a channel that can be used for communication between two nodes that are connected via this edge. This communication can be done by passing special messages between these two nodes. Indeed, messages are multivariate functions in $R$ in this message passing algorithm. Moreover, each processor (node) obeys a simple rule for generating the ongoing message from a specific edge that is connected to this node. This rule, which is also known as the sum-product update rule, is as following:
\begin{center}
\begin{minipage}{16cm}
\begin{center}
\begin{minipage}{15.1cm}
\vspace*{1mm}
\textbf{The Sum-Product Update Rule}~\cite{frey1997factor,kschischang2001factor}: The messages sent from a node $v$ to a node $u$ on an edge $e$ is the product of the local function at $v$ (if any) with all messages received at $v$ on edges \textit{other} than $e$, summarized for the variable associated with $e$, i.e. $x_{u,v}$.
\vspace*{1mm}
\end{minipage}
\end{center}
\end{minipage}
\end{center}

It is important to mention that by summarizing the product of incoming messages with respect to the variable node associated with an edge, the messages between a function node ($f$) and a variable node ($x$) is a function of just $x$. So, one can denote the message that is sent from node $v$ to node $u$ as $\mu_{v\rightarrow u}(x_{\{v,u\}})$. Moreover, if we denote the neighbours of a node in a graph by $n(.)$, then by using this notation we can describe the sum product update rule in a mathematical way as following~\cite{kschischang2001factor}:
\begin{itemize}
\item The message sent from variable node $x$ to function node $f$: 
\begin{equation}
\label{eq16}
\mu_{x\rightarrow f}(x)=\prod_{h\in n(x)\backslash \{f\}}\mu_{h\rightarrow x}(x)~~.
\end{equation}
\item The message sent from function node $f$ to variable node $x$:
\begin{equation}
\label{eq17}
\mu_{f\rightarrow x}(x)=\left ( f(X_{n(f)})\prod_{y\in n(f)\backslash \{x\}}\mu_{y\rightarrow f}(y)
  \right )\downarrow x~~.
\end{equation}
\end{itemize} 

Another point to mention is that one can find different variants of a sum-product algorithm by using appropriate definitions for $R$ and ``$+$" and ``$.$" operators, so that $(R,+,.)$ would be a semi-ring. Indeed, by using the semi-ring of $(\mathbb{R},\textrm{min},+)$ one can find an algorithm similar to sum-product algorithm which is known as min-sum algorithm. This algorithm can perform on factor graph representations of multivariate functions that are decomposed as sums of local functions. Moreover, it can find an algorithm for minimizing these types of functions over all configurations of a subset of variables. This method will be used in the later parts of this work.

Since the message that is sent from one node to another generally depends on all incoming message at this node, an initial point for sending the messages, as well as a certain time scheduling, are required. For the initial state, it can be assumed that a unit message (a message corresponding to unit product function) is received at all of incoming lines of each node before this initial state. Hence, every node is ready to send its message at the initial state. For the scheduling, it is assumed that we have an external clock, with which the transmission of messages is synchronized~\cite{kschischang2001factor,aji2000generalized}. Further, only one message at each direction can be transmitted over an edge at each clock pulse. This message, will take the place of the previous message. As a rule, a message that is sent at time $t=i$ from node $v$ only depends on the local function at $v$ and on the newest messages that are arrived before $i$. For scheduling, there are a number of methods such as flooding schedule and serial schedule~\cite{aji2000generalized}. However, for a factor graph that does not have any loops (i.e. it is a tree), there is a famous scheduling known as Generalized Forward-Backward (GFB) that was introduced by Frey et al. in 1997~\cite{frey1997factor}. During this scheduling, only one message in total will be sent over a specific edge and at each direction. More accurately, a node $v$ that transmits messages on edge $e$ will send its message on $e$ once and for all, and will do this when it received all messages from all other edges than $e$. Initiation of this scheduling is from leaves of the tree. At the initial state, for every leaf variable node $x$ and leaf function node $f$ we have:
\begin{equation}
\label{eq18}
\mu_{x\rightarrow n(x)}(x)=1~~~~~~,~~~~~~\mu_{f\rightarrow n(f)}(X_{n(f)})=f(X_{n(f)})\enspace.
\end{equation}
By passing time after initiation at leaves, more nodes at higher hierarchical levels will became ready for sending messages~\cite{kschischang2001factor}. Hence, messages will be spread out throughout the tree to all of the nodes. Termination of the schedule will happen when all of the nodes of the tree have transmitted and received messages, to and from all of their connected edges, i.e. when we have transmitted messages in both directions at each edge.
  
The importance of GFB scheduling is due to its prominent role in computing marginal functions by the sum-product algorithm. More mathematically, we have the following theorem in~\cite{kschischang2001factor}:
\begin{theorem}
Suppose that $(R,+,.)$ is a semi-ring. Furthermore, suppose that we have a multivariate function $g:A_S\rightarrow R$, with a factor graph representation of $F(S,Q)$. Additionally, assume that this factor graph is a finite tree. If $x$ is a variable node of $F$, and if for every $f\in n(x)$, $\mu_{f\rightarrow x}$ is the message that is transmitted from $f$ to $x$ during the sum-product algorithm, and finally if we use GFB scheduling for passing the messages in the graph, then at the termination state of GFB we will have:
\begin{equation}
\label{eq19}
g(X_S)\downarrow x=\prod_{f\in n(x)}{\mu_{f\rightarrow x}(x)}\enspace.
\end{equation}  
\label{theorem1}
\end{theorem}

We can apply the above theorem to the semi-ring $(\mathbb{R},\textrm{min},+)$). Then, using min-sum algorithm instead of sum-product and then again GFB scheduling, we have the following at the termination point:
\begin{equation}
\label{eq20}
g(X_S)\downarrow x\triangleq \mathop{\textrm{{min}}}_{X_{S\backslash \{x\}}\in A_{S\backslash \{x\}}} g(X_S)=
\sum_{f\in n(x)}{\mu_{f\rightarrow x}(x)}\enspace.
\end{equation}
Using (\ref{eq20}), we can introduce a method for finding the exact solution of (\ref{eq9}), as will be discussed in the following section. 

\section{\label{sec3} Factor Graph Model and Exact Map Estimator}
In this section, we will try to apply the method of factor graph to the problem of sparse channel estimation. To do this, we will first find a factor graph representation of the cost function in (\ref{eq9}), and then will simplify it for running a message passing algorithm like min-sum algorithm. Afterwards, by applying this algorithm we will find an exact solution to the mentioned discrete optimization problem in (\ref{eq9}). It is important to mention that the direct method for finding $\mathbf{\hat{b}_{\textrm{{MAP}}}}$ involves a combinatorial search among $2^M$ possible values for $\mathbf{b}$, and so its complexity will grow exponentially with respect to channel memory ($M$). However, our proposed method, which will be introduced in this section, will find this exact solution without a need for a combinatorial search. Instead, by using a semi Viterbi algorithm based on the min-sum algorithm, we will find a solution for (\ref{eq9}), while its complexity will grow polynomially with respect to channel memory and will grow exponentially with respect to the length of the training sequence ($L$). This improvement will allow us to use our algorithm in estimation of location of non-zero taps in the case of a sparse channel whose memory is so large, by using an appropriate training sequence with an appropriate length. 

It is worth mentioning that there is a correspondence between the length of the training sequence and the Cram\'{e}r-Rao bound of the estimation. Indeed, when we are using a random Bernoulli training sequence with i.i.d elements and a fixed sum of energy among its elements, the Cram\'{e}r-Rao bound of the estimation will become higher by a decrease in the length of the training data~\cite{sharp2008estimation,niazadeh2010alternating}. Accordingly, we should make a trade-off between the computational complexity and the bound of estimation error when using our algorithm. The exact steps and derivation of our algorithm will be discussed in the following. 

\subsection{Our Proposed Factor Graph Representation}
To make a factor graph model of the MAP estimation problem in (\ref{eq9}), suppose that we use the semi-ring $(\mathbb{R},\textrm{min},+)$. Indeed, we will use this semi-ring to propose a factor graph model for the cost function in (\ref{eq9}), and to introduce a sum-product algorithm for finding the solution of (\ref{eq9}), which in this case would be called a min-sum algorithm. To do this, we can rewrite the discrete optimization problem over $\{0,1\}^M$ in (\ref{eq9}) as following:
\begin{align}
\label{eq21}
\mathbf{\hat{b}}_{\text{{MAP}}}&=
\underset{\mathbf{\tilde{b}}\in\{0,1\}^M }{\textrm{{argmin}}}~\lVert\mathbf{y}-\mathbf{U_{\hat{h}}}\mathbf{\tilde{b}}\rVert_2^2+\lambda \lVert \mathbf{\tilde{b}}\rVert_0=\underset{\mathbf{\tilde{b}}\in\{0,1\}^M }{\textrm{{argmin}}}~
\mathbf{\tilde{b}}^T\mathbf{U_{\hat{h}}}^T\mathbf{U_{\hat{h}}}\mathbf{\tilde{b}}-2\mathbf{y}^T\mathbf{U_{\hat{h}}}\mathbf{\tilde{b}}
+\lambda\sum_{i=0}^{M-1}{\tilde{b}_i}\enspace.
\end{align}
Furthermore, by defining $\mathbf{X}_{M\times M}\triangleq \mathbf{U_{\hat{h}}}^T\mathbf{U_{\hat{h}}}$ and $\mathbf{z}_{M\times 1}\triangleq\mathbf{U_{\hat{h}}}^T\mathbf{y}$, we can rewrite the cost function in (\ref{eq21}) as:
\begin{align}
\label{eq22}
\sum_{i=0}^{M-1}{\sum_{j=0}^{M-1}{\tilde{b}_i\tilde{b}_j X_{i,j}}}-\sum_{i=0}^{M-1}2z_i\tilde{b}_i+\lambda\sum_{i=0}^{M-1}{\tilde{b}_i}\enspace.
\end{align}
By assuming $L\leq M$ (in which $L$ is the length of training sequence and $M$ is the channel memory), $\mathbf{U}$ will be in the form described in (\ref{eq3}). Additionally, $\mathbf{U_{\hat{h}}}=
\mathbf{U}\textrm{diag}(\mathbf{\hat{h}})$ could be obtained by multiplying the $i$-th column of $\mathbf{U}$ by $\hat{h}_i$ for every $0\leq i\leq M-1$. Consequently, according to the form of matrix $\mathbf{U}$ in (\ref{eq3}) and by the expression $\mathbf{X}=\mathbf{U_{\hat{h}}}^T\mathbf{U_{\hat{h}}}$, we have the following property for every matrix $\mathbf{X}$:
\begin{align}
\label{eq23}
&\lvert i-j \rvert\geq L:~~X_{i,j}=\text{$<$ $i$-th column of $\mathbf{U_{\hat{h}}}$~,~$j$-th column of $\mathbf{U_{\hat{h}}}$ $>$}=0\enspace,
\end{align}
in which $<.>$ denotes the inner product operator. Additionally, $X$ is a symmetric matrix. Accordingly, by the use of (\ref{eq23}), the cost function in (\ref{eq22}) can be rewritten as:
\begin{align}
\label{eq24}
g(\mathbf{\tilde{b}})&\triangleq
\sum_{i=0}^{M-1}{\sum_{j=i-L+1}^{i-1}{2\tilde{b}_i\tilde{b}_j X_{i,j}}}-\sum_{i=0}^{M-1}2z_i\tilde{b}_i+\lambda\sum_{i=0}^{M-1}{\tilde{b}_i}=\sum_{i=0}^{M-1}\tilde{b}_i[\tilde{b}_i X_{i,i}+\sum_{j=i-L+1}^{i-1}{2\tilde{b}_j X_{i,j}}-2z_i+\lambda].
\end{align}
Now, by defining $M$ local functions $\tilde{f}_i(\tilde{b}_i,\tilde{b}_{i-1},\dots,\tilde{b}_{i-L+1})\triangleq \tilde{b}_i[\tilde{b}_i X_{i,i}+\sum_{j=i-L+1}^{i-1}{2\tilde{b}_j X_{i,j}}-2z_i+\lambda]$ for every $0\leq i\leq M-1$, we have an additive decomposition of $g(\mathbf{\tilde{b}})$ to local functions $\{\tilde{f}_i: 0\leq i\leq M-1\}$, i.e. $g(\mathbf{\tilde{b}})=\sum_{i=0}^{M-1}\tilde{f}_i$. Using the semi-ring $(\mathbb{R},\textrm{min},+)$, the factor graph representation of $g(\mathbf{\tilde{b}})$ will be as described in the Fig.~\ref{fig1}-a.

Another point to mention is that in (\ref{eq24}), according to having the observation vector $\mathbf{y}$ at the receiver side, the elements of this vector are not considered as variable nodes. More accurately, by considering $\mathbf{y}$ as a vector of variables, we will have a cost function in the form of $f(\mathbf{\tilde{b}},\mathbf{y})$. In this case, $g(\mathbf{\tilde{b}})$ will be the cross section of $f$ with respect to $\mathbf{y}$, as was defined in (\ref{eq12}). So, $g(\mathbf{\tilde{b}})=f(\mathbf{\tilde{b}}||\mathbf{y})$ and the associated factor graph representation (in which each $\tilde{f}_i$ is a function of $\{y_i\}_{i=0}^{M-1}$) is as described in Fig.~\ref{fig1}-b. Using the method of finding the factor graph representation of a cross section (Sec.~\ref{sec2}), it is almost obvious that the factor graph in Fig.~\ref{fig1}-a is indeed the corresponding cross section factor graph representation with respect to $\mathbf{y}$.

As it can be seen from Fig.~\ref{fig1}-a, the factor graph of our cost function has several loops and it is not a tree. Hence, Theorem.~\ref{theorem1} cannot be applied directly to this factor graph. In the following section, we will provide an equivalent factor graph representation for $g$, in which there is no loop in it, and the min-sum algorithm can be performed directly using the GFB scheduling (Theorem~\ref{theorem1}).
\subsection{Simplifying The factor graph For Running Min-Sum Algorithm}
\begin{figure}[h]   
    \graphicspath{{pics/}}
    \centering		
	\subfigure[]{
    \includegraphics[scale=0.52]{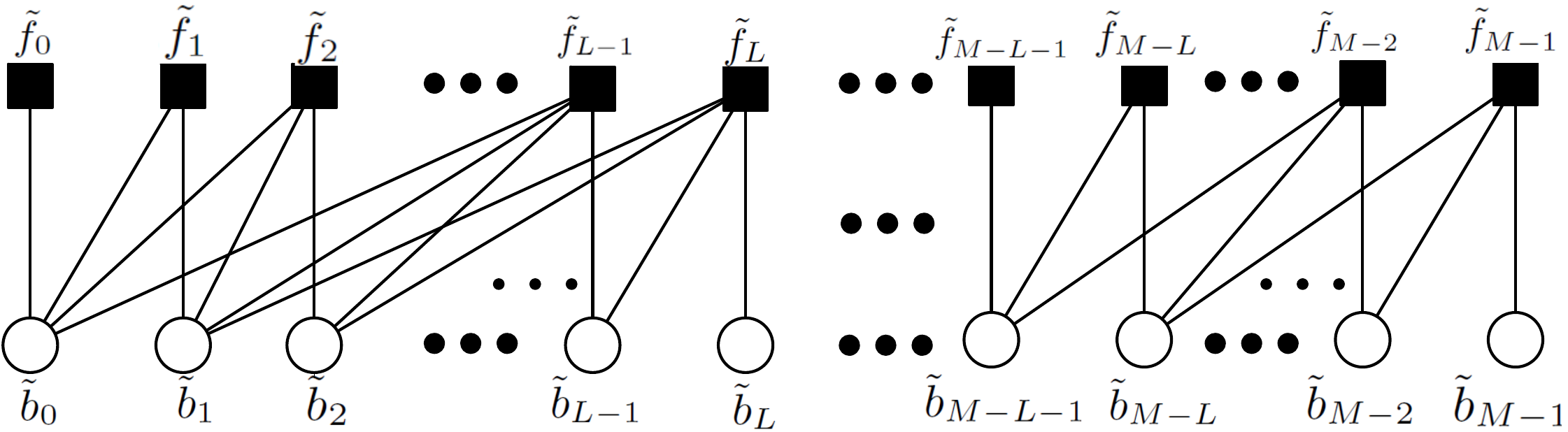}
	}	
	\subfigure[]{
    \includegraphics[scale=0.52]{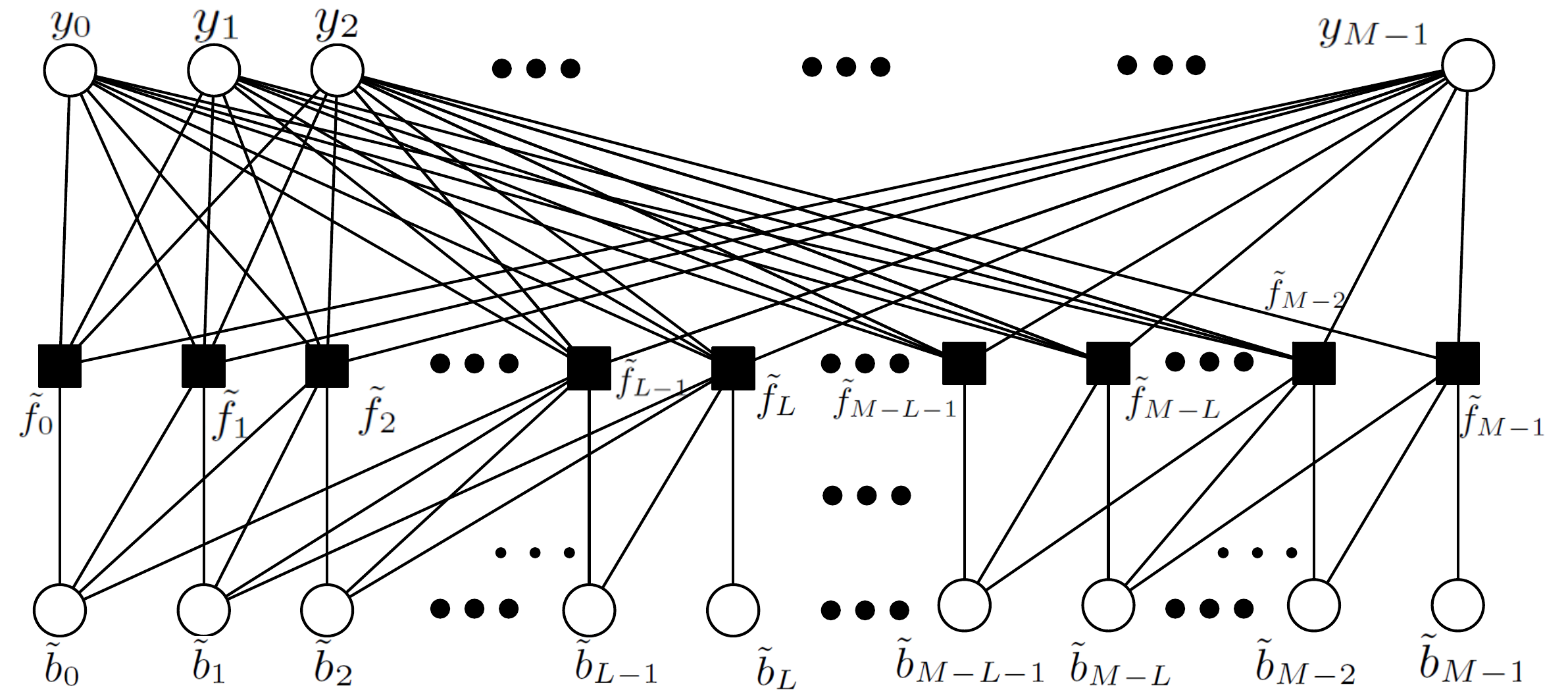}
	}	
	\caption{\label{fig1} Factor graph representation of (a) $g(\mathbf{\tilde{b}})=f(\mathbf{\tilde{b}}||\mathbf{y})$ and of (b) $f(\mathbf{\tilde{b}},\mathbf{y})$.}    
\end{figure}

For removing the mentioned loops of the factor graph in Fig.~\ref{fig1}-a, firstly we define $M+L+1$ latent variables $\mathbf{s}_i\triangleq(\tilde{b}_{i-1},\tilde{b}_{i-2},\dots,\tilde{b}_{i-L+1})$ for every $0\leq i\leq M+L$ which we call them \textit{state variables}. More accurately, we have:
\begin{equation}
\label{eq25}
\mathbf{s}_i \triangleq \left\{
\begin{array}{ll}
0                ~~~& i=0\\
(\tilde{b}_{i-1},\tilde{b}_{i-2},\dots,\tilde{b}_{0})    ~~~& 1\leq i\leq L-1\\
(\tilde{b}_{i-1},\tilde{b}_{i-2},\dots,\tilde{b}_{i-L+1})~~~& L\leq i\leq M\\
(\tilde{b}_{M-1},\tilde{b}_{M-2},\dots,\tilde{b}_{i-L+1})~~~& M+1\leq i\leq M+L-1\\
0              ~~~&i=M+L
\end{array} \right. 
\end{equation}
\begin{figure}[htb]   
    \graphicspath{{pics/}}
    \centering
    \includegraphics[scale=0.35]{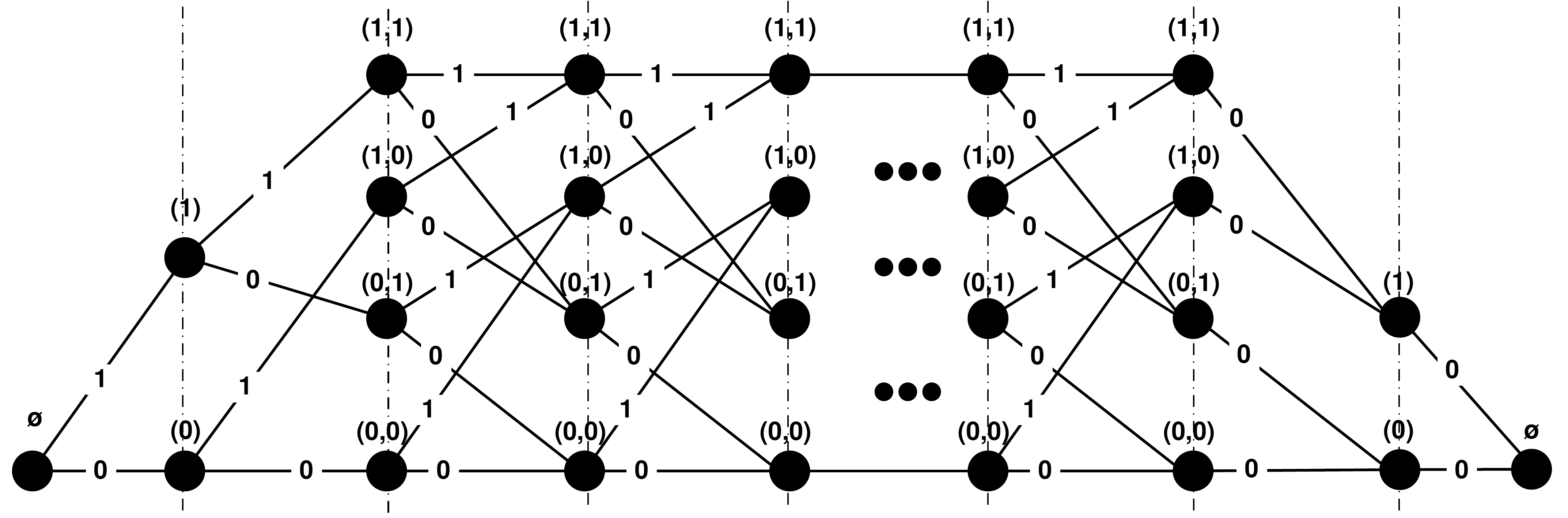}
	\caption{\label{fig2} trellis diagram for variables $\{\mathbf{s}_i\}_{i=0}^{L+M}$ and $\{\tilde{b}_i\}_{i=0}^{M-1}$}    
\end{figure}

Additionally, we will define $\mathbf{S}\triangleq(\mathbf{s}_0,\mathbf{s}_1,\dots,\mathbf{s}_{M+L})$ as the states space. It can be seen from the above definition that by knowing $\mathbf{s}_{i-1}$ and $\tilde{b}_i$ one can determine $\mathbf{s}_i$. In other words, the relationship between variables $\{\mathbf{s}_i\}_{i=0}^{L+M}$ and $\{\tilde{b}_i\}_{i=0}^{M-1}$ can be described in a trellis diagram, similar to the trellis diagrams that appear in decoding of convolutional codes in coding theory~\cite{lin1983error}, or in sequence-by-sequence maximum likelihood equalization for ISI channels in the context of digital communication~\cite{proakis2001digital}. An example of such a digram for $L=3$ and $M=5$ can be seen in Fig.~\ref{fig2}. In this diagram, nodes and edges indicate the possible values for state variables and possible values for $\{\tilde{b}_i\}_{i=0}^{M-1}$ respectively. By increasing $i$ from $0$ to $L+M-1$, we move from the left of the diagram to the right of it. The $i$-th section of the digram from left corresponds to the variable $\tilde{b}_i$ and these sections have been separated by vertical lines in Fig.~\ref{fig2}. For convenience, we also define $\mathbf{T}_i$ as the set of all possible values for triplet $(\mathbf{s}_{i},\tilde{b}_{i},\mathbf{s}_{i+1})$, i.e. all triplets $(\mathbf{s}_{i},\tilde{b}_i,\mathbf{s}_{i+1})$ that satisfy the following equation:
\begin{equation}
\label{eq26}
 \mathbf{s}_{i+1}=(\tilde{b}_{i}, \mathbf{s}_{i}^{(1)})\enspace, 
\end{equation}
in which $ \mathbf{s}_{i}^{(1)}$ is a vector that can be extracted from $\mathbf{s}_{i}$  by removing the element $\tilde{b}_{i-L+1}$. 

Similar to (\ref{eq11}), we would also use a new definition for $[P]$ in the semi-ring $(\mathbb{R},\textrm{{min}},+)$ as following:
\begin{equation}
\label{eq27}
[P] = \left\{
\begin{array}{rl}
0& ~~~~~ \text{if $P$, } \\
+\infty& ~~ \text{if not $P$. }
\end{array} \right.
\end{equation}
Note that in this mentioned semi-ring, $0$ is the identity element for ``$+$" operator ($\forall x\in \mathbb{R}: x+0=0$) and $+\infty$ is the identity element for ``$\textrm{min}$" operator ($\forall x\in \mathbb{R}:\textrm{{min}}(x,+\infty)=x$). Hence, we can conclude that (\ref{eq27}) is just the equivalent form of (\ref{eq11}) in semi-ring $(\mathbb{R},\textrm{{min}},+)$.

\begin{figure}[htb]   
    \graphicspath{{pics/}}
    \centering
    \includegraphics[scale=0.52]{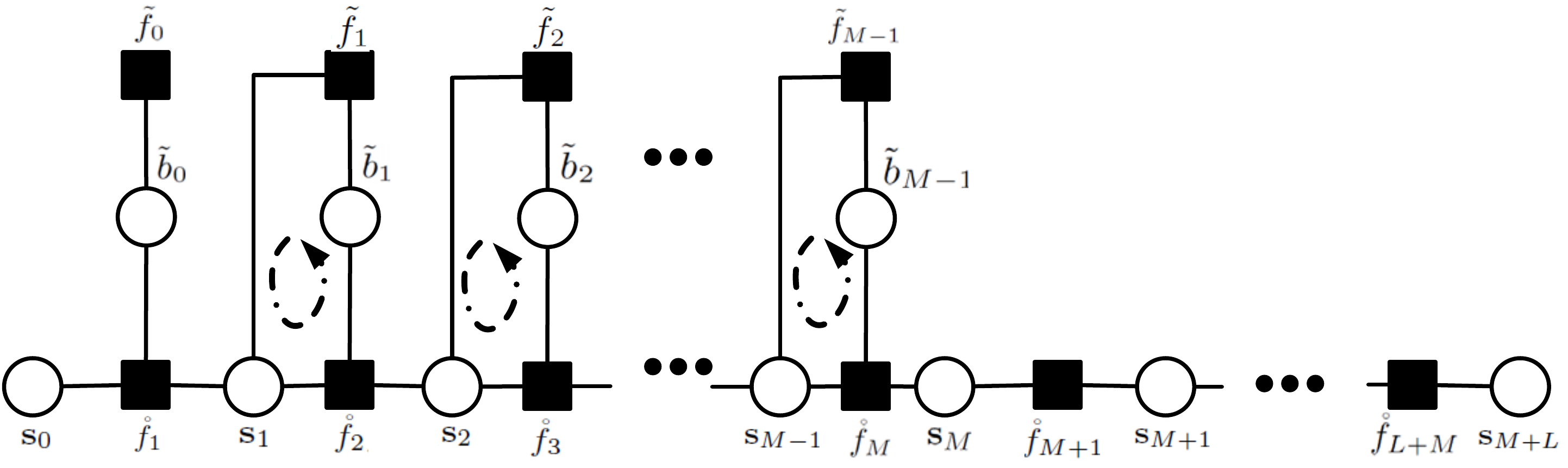}
	\caption{\label{fig3} Equivalent factor graph representation after introducing the state variables}    
\end{figure}

Now, by using the definition in (\ref{eq27}) and the fact that for every $0\leq i\leq M-1$, $f_i$ is just a function of $\tilde{b}_i$ and $\mathbf{s}_i$, we have the following equivalent form for the optimization problem in (\ref{eq24}):
\begin{align}
\underset{\mathbf{\tilde{b}}\in\{0,1\}^M }{\textrm{{argmin}}}~g(\mathbf{\tilde{b}})&=
\underset{\mathbf{\tilde{b}},\mathbf{S}:(\mathbf{s}_{i-1},\tilde{b}_i,\mathbf{s}_i)\in \mathbf{T}_i,0\leq i\leq L+M}{\textrm{{argmin}}}~\sum_{i=0}^{M-1}{\tilde{f}_i(\tilde{b}_i,\mathbf{s}_i)}\nonumber\\
&\label{eq28}=\underset{\mathbf{\tilde{b}},\mathbf{S}}{\textrm{{argmin}}}~ \{\sum_{i=0}^{M-1}{\tilde{f}_i(\tilde{b}_i,\mathbf{s}_i)}\nonumber\\
&+\sum_{i=0}^{L+M-1}{[(\mathbf{s}_{i},\tilde{b}_i,\mathbf{s}_{i+1})\in \mathbf{T}_i]}\}\enspace.
\end{align}
In (\ref{eq28}), the second equality comes from adding the terms  $[(\mathbf{s}_{i},\tilde{b}_i,\mathbf{s}_{i+1})\in \mathbf{T}_i]$ and removing the constraints on feasible region, which can be done according to the definition (\ref{eq27}) for $[.]$. By defining $\tilde{g}(\mathbf{S},\mathbf{\tilde{b}})\triangleq\sum_{i=0}^{M-1}{\tilde{f}_i(\tilde{b}_i,\mathbf{s}_i)}+\sum_{i=0}^{L+M-1}{[(\mathbf{s}_{i},\tilde{b}_i,\mathbf{s}_{i+1})\in \mathbf{T}_i]}$ and $\mathring{f}_i\triangleq[(\mathbf{s}_{i-1},\tilde{b}_{i-1},\mathbf{s}_{i})\in \mathbf{T}_{i-1}]$, the factor graph representation of $\tilde{g}$ will be as described in Fig.~\ref{fig3}. As it can be seen, this factor graph is not yet a tree, and indeed has several loops (which are indicated in Fig.~\ref{fig3}). To omit these loops, we will use the following theorem:

\begin{theorem}[\label{theorem2}Equivalent Spanning Tree~\cite{kschischang2001factor}]
Suppose that $F$ is a factor graph and $T$ is a spanning tree of $F$, i.e. a sub-tree of $F$ that passes from all nodes of $F$. Now, suppose that $x$ is a variable node of $F$ and one of the edges connected to this node (for example $\{x,f\}$) does not belong to $T$. So, there is a loop in $F$ that includes both the edge $\{x,f\}$ and a unique path  from $x$  to $f$ in $T$. We call this loop $C$. In this case, if we \textit{distribute} the variable $x$ in all of the variable nodes in $C$ (i.e. if  we replace each variable node $v\in C$ by $(v,x)$) and delete the edge $\{x,f\}$ from $F$, then we will obtain a factor graph $F'$ which is \textit{equivalent} with $F$. By equivalent, we mean that the results of sum-product algorithm (and obviously its variants) are the same in both of these graphs.
\end{theorem}

\begin{figure}[h]   
    \graphicspath{{pics/}}
    \centering
    \includegraphics[scale=0.52]{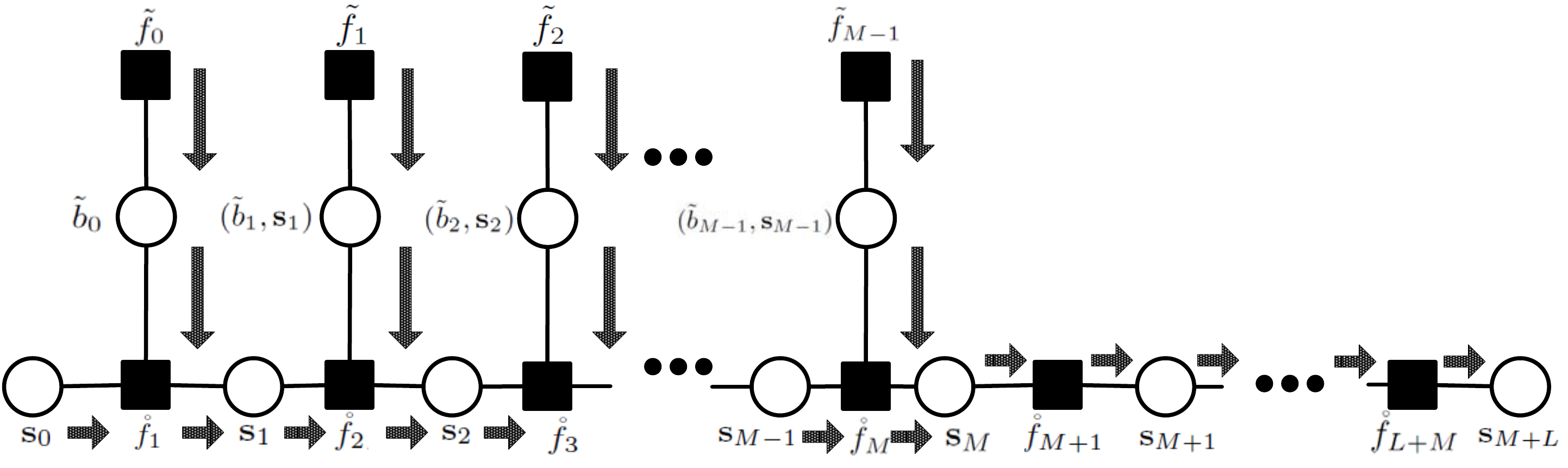}
	\caption{\label{fig4} Final factor graph}    
\end{figure}

By applying Theorem.~\ref{theorem2} on the factor graph of Fig.~\ref{fig3}, we will first distribute the variables nodes $\{\mathbf{s}_i\}_{i=1}^{M-1}$ in the remaining loops of the graph (that are indicated in Fig.~\ref{fig3}), and then we will remove the remaining edges that do not belong to the spanning tree. Thus, for every $1\leq i\leq M-1$ we will replace the node $\tilde{b}_i$ by $(\tilde{b}_i,\mathbf{s}_i)$, and then will remove the edge $\{\mathbf{s}_i,\tilde{f}_i\}$. The resulted factor graph, which is indeed a tree, is described in Fig.~\ref{fig4}. Using this graph, we can perform the min-sum algorithm with GFB scheduling, which will be discussed in the following sub-section.
\subsection{Implementation of Min-Sum Algorithm On a Trellis Diagram}

Consider the multivariate function $\tilde{g}(\mathbf{S},\mathbf{\tilde{b}})$. As it can be seen from (\ref{eq28}), and according to the definition of $\downarrow$ operator in the semi-ring $(\mathbb{R},\textrm{{min}},+)$, we are looking forward to compute $\tilde{g}(\mathbf{S},\mathbf{\tilde{b}})\downarrow X_{\emptyset}$, in which $X_\emptyset$ is the empty set of variables. Further, by knowing that we have just one possible value for $\mathbf{s}_{L+M}$ (i.e. $\mathbf{s}_{L+M}=0$), we can say that we are looking forward to find $\tilde{g}(\mathbf{S},\mathbf{\tilde{b}})\downarrow \mathbf{s}_{L+M}$. As was mentioned before, we can find a way for computing this expression using the min-sum algorithm that is run under GFB scheduling. More accurately, if we perform min-sum algorithm with GFB scheduling on the factor graph representation of
$\tilde{g}(\mathbf{S},\mathbf{\tilde{b}})$ (which is indeed a tree), and compute the message that is received at $\mathbf{s}_{L+M}$, then we can find the solution to the problem in (\ref{eq28}). As it can be seen from Fig.~\ref{fig4}, this message is produced by a flow of message passed towards $\mathbf{s}_{L+M}$ through a direct path that is also indicated in Fig.~\ref{fig4}.

By defining $e=(\mathbf{s}_i,\tilde{b}_i,\mathbf{s}_{i+1})$ in the mentioned trellis diagram (such as the one in Fig.~\ref{fig2}), as well as defining $\mu_{\mathring{f}_{i+1}\rightarrow \mathbf{s}_{i+1}}\triangleq\alpha(\mathbf{s}_{i+1})$,
$\mu_{\mathring{f}_{i}\rightarrow \mathbf{s}_{i}}=
\mu_{\mathbf{s}_{i}\rightarrow \mathring{f}_{i+1}}\triangleq \alpha(\mathbf{s}_i)$ and $\mu_{\tilde{f}_i\rightarrow (\tilde{b}_i,\mathbf{s}_i)}=
\mu_{(\tilde{b}_i,\mathbf{s}_i)\rightarrow \mathring{f}_i}=\tilde{f}_i(\mathbf{s}_i,\tilde{b}_i)
\triangleq \gamma (e)$, we can say that in the $i$-th section of the min-sum algorithm ($0\leq i\leq M-1$), we have the following equation:
\begin{align}
\label{eq29}
\alpha(\mathbf{s}_{i+1})&=\underset{\mathbf{s}_i,\tilde{b}_i}{\textrm{{min}}}~
(\alpha(\mathbf{s}_i)+\gamma(e))+[(\mathbf{s}_i,\tilde{b}_i,\mathbf{s}_{i+1})\in \mathbf{T}_i]\nonumber\\&=
\underset{(\mathbf{s}_i,\tilde{b}_i,\mathbf{s}_{i+1})\in \mathbf{T}_i}{\textrm{{min}}}~(\alpha(\mathbf{s}_i)+\gamma(e))\enspace.
\end{align}

\begin{figure}[h]   
    \graphicspath{{pics/}}
    \centering
    \includegraphics[scale=0.52]{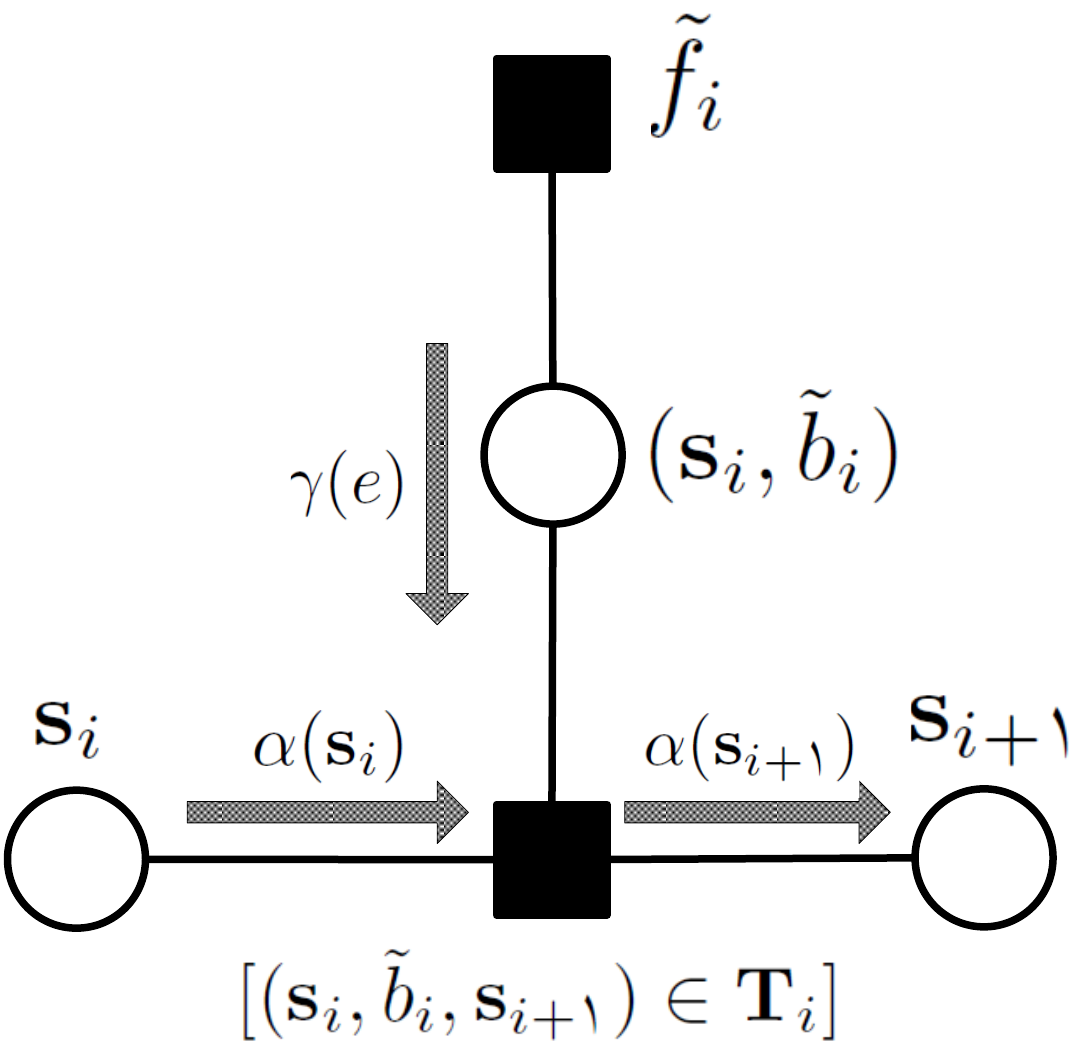}
	\caption{\label{fig6} $i$-th stage of min-sum algorithm in our factor graph }    
\end{figure}

This equation in (\ref{eq29}) can also be seen in Fig.~\ref{fig6}. Using this figure, we may see $\alpha(\mathbf{s}_i)$ as the corresponding weight for a specific value of state variable $\mathbf{s}_i$, and $\gamma(e)$ as the weight of $e$ in the trellis diagram described in previous sections. Accordingly, at each stage, the weight of each value of the state variable at this stage will be the minimum among all possible sum of $\alpha(\mathbf{s}_i)$ and $\gamma(e)$ (sum of the edge weight and weight of a specific value of the previous state variable that is connected via this edge in the trellis diagram to our value at current stage). In short, by doing a simple procedure that involves ``addition" , ``comparison" and ``selection" at each stage, we can find the appropriate weight of the state variable at each stage, for different values of this state variable. Consequently, this procedure will finally result in the weight of the only value of the state variable $\mathbf{s}_{L+M}$ (``$0$" ), and this weight will be indeed the solution of $\underset{\mathbf{S},\mathbf{\tilde{b}}}{\textrm{{min}}}~\tilde{g}(\mathbf{S},\mathbf{\tilde{b}})$.

It is important to mention that our proposed algorithm can be implemented completely on a trellis diagram, and in a way similar to the ordinary Viterbi algorithm. More accurately, we can choose some selective paths at each stage of our algorithm in the trellis diagram described before, similar to the selection of survivor paths in the context of the Viterbi algorithm~\cite{lin1983error,proakis2001digital}. Indeed, we need to select a survivor path in correspondence to each state value at every stage, and hence we need to save $2^{L-1}$ paths at each stage\footnote{Accordingly, the computational complexity of the final estimator will increase exponentially with respect to the length of training sequence and will increase linearly with respect to the channel memory.}. In mathematical words, if we define $m_i(x)$ as one of $2^{L-1}$ selected (survived) paths at $i$-th stage, i.e. the path that corresponds to the value of $x$ for state variable $\mathbf{s}_i$, then the next stage survivor paths will be updated using a simple rule: Assuming that $e^*=(\mathbf{s}^*_i,\tilde{b}^*_i,\mathbf{s}_{i+1})$ is a solution of (\ref{eq29}), then we have: 
\begin{equation}
\label{eq30}
m_{i+1}(\mathbf{s}_{i+1})=m_{i}(\mathbf{s}^*_i)|\tilde{b}^*_i\enspace,
\end{equation}
in which $|$ denotes the operator of concatenation of two paths. At the end of the algorithm, $m_{L+M}(0)$ will determine  $\mathbf{\hat{b}}_{\textrm{{MAP}}}$. It is worth to mention that our 
procedure is very similar to the Viterbi algorithm, but the weight of each branch (edge) of the trellis diagram is determined by the value of state variables at both ends of this edge. 

In conclusion, we will use the proposed exact estimator in finding the solution of (\ref{eq9}). Afterwards, we will use it as the section for MAP estimation of locations of non-zero taps in our alternating minimization algorithm in~\cite{niazadeh2010alternating}, and finally we will introduce an algorithm for sparse channel estimation. The performance and estimation accuracy of this algorithm, which we call it Optimal MAP estimator based on Factor Graph (OMAPFG), will be investigated in Sec.~\ref{sec4}. Additionally, we will compare the MSE of our proposed algorithm with CRB-S, and will show by the use of simulation that our algorithm can perform very close to CRB-S, even at low SNR. 
\section{\label{sec4} Experimental Results}
\begin{figure}[h]   
    \graphicspath{{pics/}}
    \centering
    \includegraphics[scale=0.52]{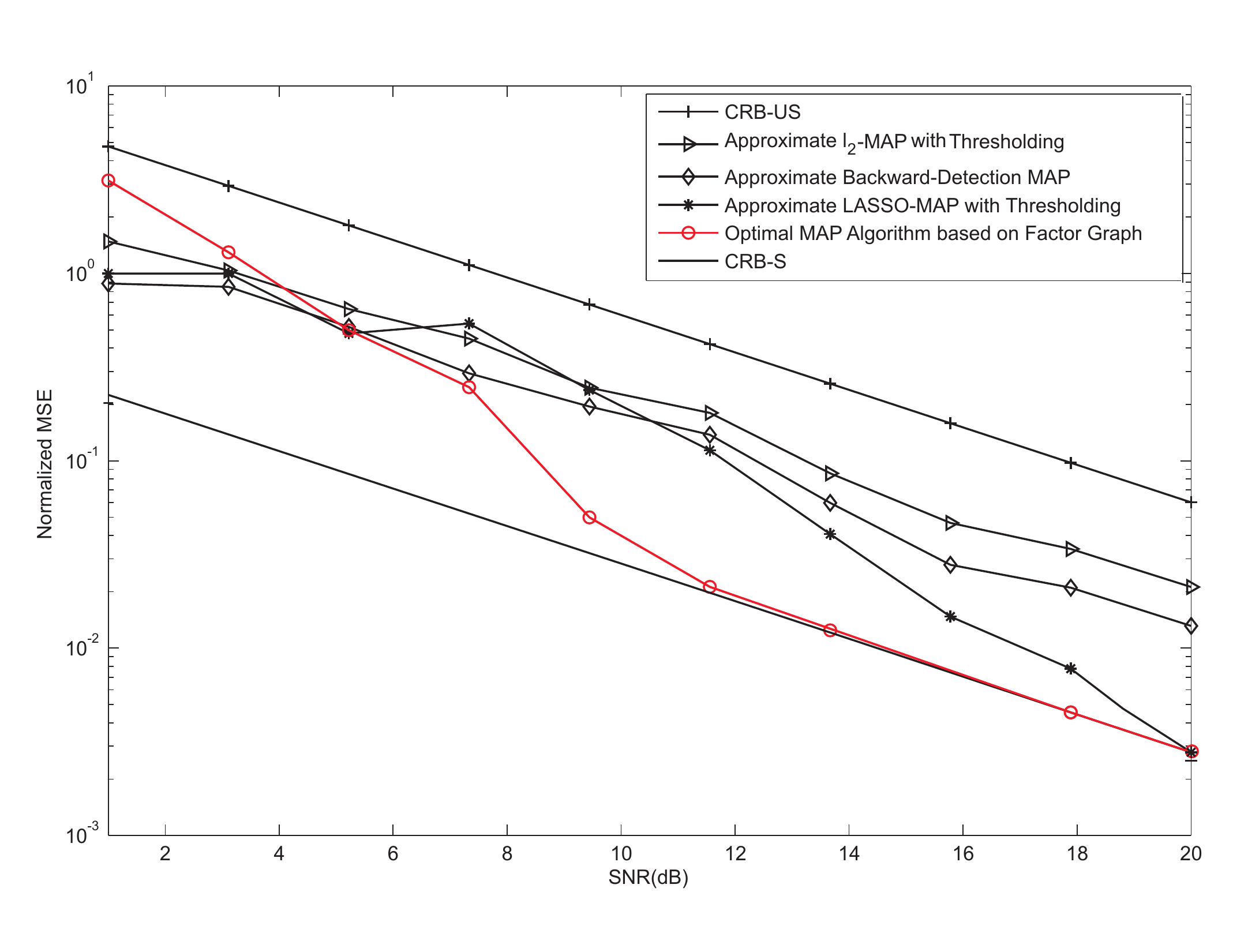}
	\caption{\label{fig7} Comparison of OMAPFG with other variants of method of alternating minimization}    
\end{figure}

\begin{figure}[h]   
    \graphicspath{{pics/}}
    \centering
    \includegraphics[scale=0.52]{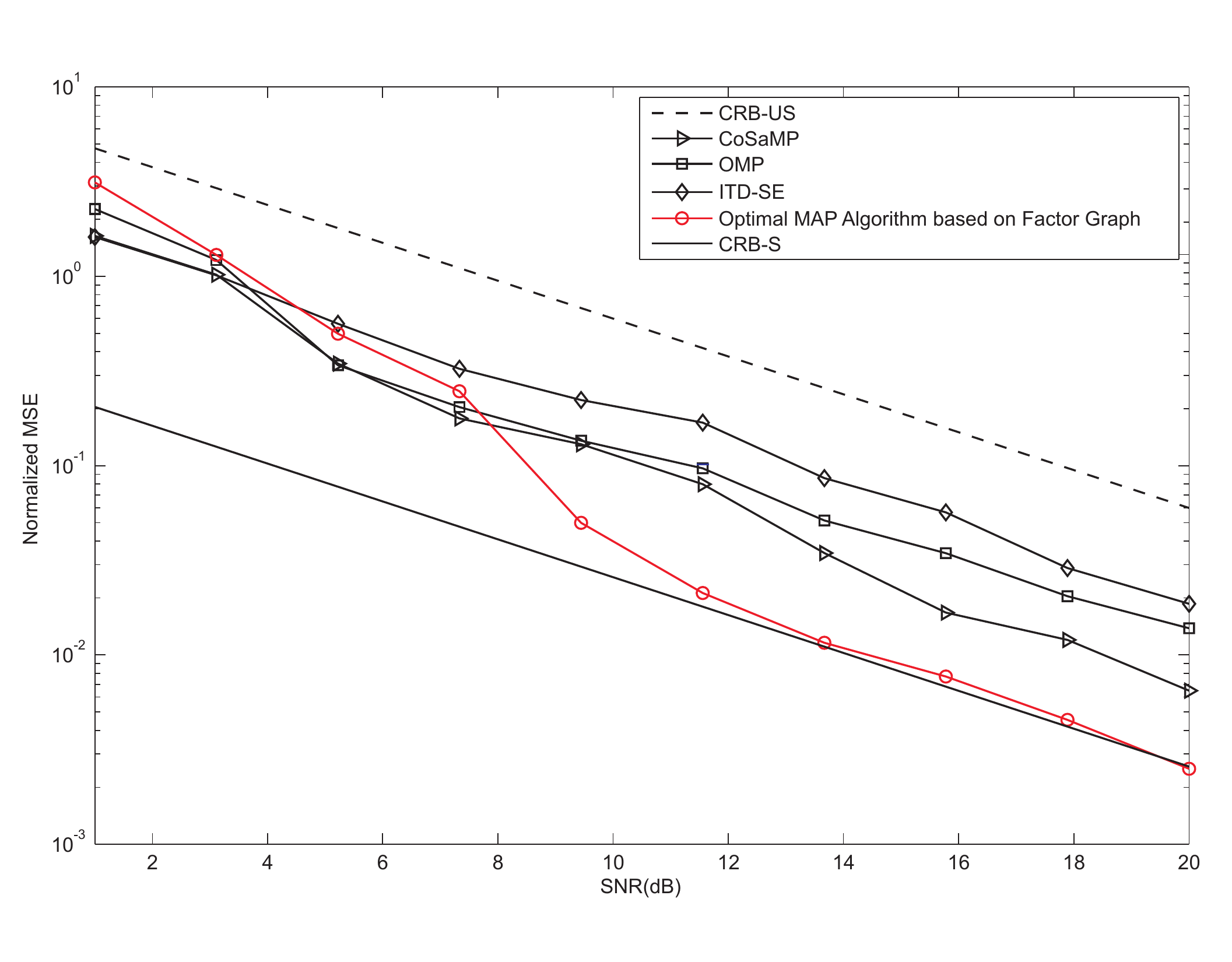}
	\caption{\label{fig8} Comparison of OMAPFG with some methods of sparse channel estimation}    
\end{figure}

In this part, we examine the MSE performance of the proposed OMAPFG algorithm in an experiment and we will compare it to ITD-SE~\cite{carbonelli2007sparse}, CoSaMP~\cite{needell2009cosamp}, OMP~\cite{karabulut2004sparse}, Cram\'{e}r-Rao Bound of the structured estimator (CRB-S), Cram\'{e}r-Rao Bound of the unstructured estimator (CRB-US), and also other variants of the method of alternating minimization that use an approximate MAP estimation for the locations of non-zero taps. In this experiment, we choose $M=30$ and $K=5$. Hence, we generate a separate $K$-sparse random signal as the sparse channel for this experiment. We will also use $L=5$ as the length of the training sequence, $\mathbf{u}$. Moreover, elements of $\mathbf{u}$ are generated independently random according to a symmetric Bernoulli distribution of symbols in $\{\pm 1\}$. For finding MSE, each experiment is repeated $100$ times and the averaging over all these runs is taken. We run the simulations on an $2.8$Ghz Intel Core2Duo CPU. The Normalized MSE vs SNR curves for all of the algorithms are computed. Comparison of the proposed OMAPFG and all of the variants of the method of alternating minimization in~\cite{niazadeh2010alternating} are shown in Fig. \ref{fig7}. Furthermore, comparison of the proposed OMAPFG, CoSaMP, OMP, and ITD-SE are shown in Fig. \ref{fig8}. In both of these figures, CRB-S and CRB-US for MSE are drawn, which are equal to $\sigma^2\mbox{Tr}\{(\mathbf{U_h}^T\mathbf{U_h})^{-1}\}$ and $\sigma^2\mbox{Tr}  \{(\mathbf{U}^T\mathbf{U})^{-1}\}$ respectively as in \cite{niazadeh2010alternating,sharp2008estimation}.
 Computational complexities of different algorithms are compared by the use of total CPU times needed for convergence in all $100$ run of the experiment, which are shown in Tab. \ref{table1}. We use CPU time as a metric for roughly comparison of computational complexity. Note that for the validity of our comparison, we run each algorithm until the condition of convergence is satisfied, i.e. when $\frac{\lVert\mathbf{\hat{h}^{(i)}}-\mathbf{\hat{h}^{(i-1)}}\rVert_2^2}{\lVert\mathbf{\hat{h}^{(i)}}\rVert_2^2}\leq \epsilon$, in which $\epsilon >0$ is an arbitrary small enough positive number (we choose  value of $0.01$ for $\epsilon$ in our simulation).
 \begin{table}[h]
\caption{\label{table1} Comparison of CPU times for different algorithms}
\centering
\begin{tabular}{||l|l||}
\hline
\multicolumn{2}{||c||}{CPU time in seconds for $100$ experiments} \\
\hline
{Optimal MAP based on Facotr Graph} & $9.52$\\
\hline
{Approximate LASSO-MAP with Thresholding} & $12.13$\\
\hline
{ITD-SE} & $8.30$\\
\hline
{OMP} & $16.20$\\
\hline
{CoSaMP} & $18.40$\\
\hline
\end{tabular}
\end{table}

As it can be seen from Fig.~\ref{fig7} and Fig.~\ref{fig8}, our proposed method of OMAPFG is almost better than other variants of alternating minimization for all SNRs in the sense of MSE. Further, by comparing OMAPFG with ITD-SE, CoSaMP, and OMP we conclude that OMAPFG has a far better accuracy in the sense of MSE in comparison to the mentioned algorithms, while also nearly achieving CRB-S even in low SNRs (around 10 dB). It is also important to mention that the direct search for finding the exact solution of (\ref{eq9}) requires an exponentially computational complexity with respect to the channel memory, while the computational complexity of OMAPFG increases linearly with respect to channel memory and increases exponentially with respect to the length of the training sequence. As a result, computational cost of implementing OMAPFG is even less than that of some other common algorithms for sparse channel estimation. More accurately, as it can be seen from Tab.~\ref{table1}, the total CPU time for OMAPFG is a little higher than ITD-SE (which is one of the fastest algorithms for sparse channel estimation~\cite{carbonelli2007sparse}), while it is less than all of other mentioned algorithms, i.e. our algorithm performs with a comparatively high speed with respect to all of the mentioned algorithm, while it also performs near CRB-S in the sense of MSE. In short, OMAPFG can be considered as a \textit{fast} and \textit{near efficient} estimator for sparse channel, that is, it can almost achieve CRB-S in middle and high SNRs in a relatively small time in comparison to other algorithms in the literature. 
\section{Conclusion}
In this paper, we provided a new method for sparse channel estimation which is based on the method of alternating minimization and uses an exact MAP estimator of locations of non-zero taps at each iteration. For the MAP estimation of locations of non-zero taps, we used the method of factor graph and message passing algorithms. Indeed, by providing a factor graph model of the MAP estimation problem, and then by using the min-sum algorithm, we found a semi-Viterbi algorithm for detecting the location of non-zero taps. Furthermore, by using simulation we showed that our proposed method is near efficient, i.e. it can nearly achieve the CRB-S in almost all middle and high SNRs, while its complexity grows polynomially with respect to channel memory. It is important to mention that we assume knowing the sparsity degree of the channel,i.e. $K$, at the receiver side. For further studies, one can investigate the behaviour and performance of our proposed estimator if the assumed value of $K$ is wrong, either too overestimated or underestimated. 
\section*{Acknowledgement}
The authors would like to thank Maryam Sharifzadeh, M.S. student of Mathematical Science Department at Sharif University of Technology, for her helpful comments. 
\bibliographystyle{IEEEtran}
\bibliography{refs}

\begin{thebibliography}{10}
\providecommand{\url}[1]{#1}
\csname url@samestyle\endcsname
\providecommand{\newblock}{\relax}
\providecommand{\bibinfo}[2]{#2}
\providecommand{\BIBentrySTDinterwordspacing}{\spaceskip=0pt\relax}
\providecommand{\BIBentryALTinterwordstretchfactor}{4}
\providecommand{\BIBentryALTinterwordspacing}{\spaceskip=\fontdimen2\font plus
\BIBentryALTinterwordstretchfactor\fontdimen3\font minus
  \fontdimen4\font\relax}
\providecommand{\BIBforeignlanguage}[2]{{%
\expandafter\ifx\csname l@#1\endcsname\relax
\typeout{** WARNING: IEEEtran.bst: No hyphenation pattern has been}%
\typeout{** loaded for the language `#1'. Using the pattern for}%
\typeout{** the default language instead.}%
\else
\language=\csname l@#1\endcsname
\fi
#2}}
\providecommand{\BIBdecl}{\relax}
\BIBdecl

\bibitem{carbonelli2007sparse}
C.~Carbonelli, S.~Vedantam, and U.~Mitra, ``{Sparse channel estimation with
  zero tap detection},'' \emph{IEEE Transactions on Wireless Communications},
  vol.~6, no.~5, pp. 1743--1763, 2007.

\bibitem{kocic1995sparse}
M.~Kocic, D.~Brady, and M.~Stojanovic, ``{Sparse equalization for real-time
  digital underwater acoustic communications},'' in \emph{Oceans Conference
  Record(IEEE)}, vol.~3, 1995, pp. 1417--1422.

\bibitem{chapman1983deconvolution}
N.~Chapman and I.~Barrodale, ``{Deconvolution of marine seismic data using the
  l1 norm},'' \emph{Geophysical Journal of the Royal Astronomical Society},
  vol.~72, no.~1, pp. 93--100, 1983.

\bibitem{sharp2008estimation}
M.~Sharp and A.~Scaglione, ``{Estimation of sparse multipath channels},'' in
  \emph{IEEE Military Communications Conference}.\hskip 1em plus 0.5em minus
  0.4em\relax San Diego, CA, 2008, pp. 1--7.

\bibitem{cotter2002sparse}
S.~Cotter and B.~Rao, ``{Sparse channel estimation via matching pursuit with
  application to equalization},'' \emph{IEEE Transactions on Communications},
  vol.~50, no.~3, pp. 374--377, 2002.

\bibitem{niazadeh2010alternating}
R.~Niazadeh, M.~Babaie-Zadeh, and C.~Jutten, ``{An alternating minimization
  method for sparse channel estimation},'' in \emph{Ninth International
  Conference on Latent Variable Analysis and Signal Seperation (LVA-ICA,
  formerly known as ICA)}.\hskip 1em plus 0.5em minus 0.4em\relax
  Springer-Verlag, Saint-Malo, France, 2010, pp. 319--327.

\bibitem{papoulis2002probability}
A.~Papoulis, S.~Pillai, and S.~Unnikrishna, \emph{{Probability, random
  variables, and stochastic processes}}.\hskip 1em plus 0.5em minus 0.4em\relax
  McGraw-Hill New York, 2002.

\bibitem{kay1998fundamentals}
S.~Kay, \emph{{Fundamentals of Statistical signal processing, Volume 2:
  Detection theory}}.\hskip 1em plus 0.5em minus 0.4em\relax Prentice Hall PTR,
  1998.

\bibitem{babadi2009asymptotic}
B.~Babadi, N.~Kalouptsidis, and V.~Tarokh, ``{Asymptotic achievability of the
  Cram{\'e}r--Rao bound for noisy compressive sampling},'' \emph{IEEE
  Transactions on Signal Processing}, vol.~57, no.~3, pp. 1233--1236, 2009.

\bibitem{candes2007dantzig}
E.~Candes and T.~Tao, ``{The Dantzig selector: Statistical estimation when p is
  much larger than n},'' \emph{Annals of Statistics}, vol.~35, no.~6, pp.
  2313--2351, 2007.

\bibitem{haupt2006signal}
J.~Haupt and R.~Nowak, ``{Signal reconstruction from noisy random
  projections},'' \emph{IEEE Transactions on Information Theory}, vol.~52,
  no.~9, pp. 4036--4048, 2006.

\bibitem{kschischang2001factor}
F.~Kschischang, B.~Frey, and H.~Loeliger, ``{Factor graphs and the sum-product
  algorithm},'' \emph{IEEE Transactions on Information Theory}, vol.~47, no.~2,
  pp. 498--519, 2001.

\bibitem{loeliger2004introduction}
H.~Loeliger, A.~Endora~Tech, and S.~Basel, ``{An introduction to factor
  graphs},'' \emph{IEEE Signal Processing Magazine}, vol.~21, no.~1, pp.
  28--41, 2004.

\bibitem{colavolpe2005application}
G.~Colavolpe and G.~Germi, ``{On the application of factor graphs and the
  sum--product algorithm to ISI channels},'' \emph{IEEE Transactions on
  Communications}, vol.~53, no.~5, 2005.

\bibitem{lin1983error}
S.~Lin and D.~Costello, \emph{{Error control coding}}.\hskip 1em plus 0.5em
  minus 0.4em\relax Prentice-Hall Englewood Cliffs, NJ, 1983.

\bibitem{bondy1976graph}
J.~Bondy and U.~Murty, \emph{{Graph theory with applications}}.\hskip 1em plus
  0.5em minus 0.4em\relax MacMillan London, 1976, vol. 290.

\bibitem{iverson1964method}
K.~Iverson, ``{A method of syntax specification},'' \emph{Communications of the
  ACM}, vol.~7, no.~10, pp. 588--589, 1964.

\bibitem{matsumura1989commutative}
H.~Matsumura and M.~Reid, \emph{{Commutative ring theory}}.\hskip 1em plus
  0.5em minus 0.4em\relax Cambridge Univ Pr, 1989.

\bibitem{aji2000generalized}
S.~Aji and R.~McEliece, ``{The generalized distributive law},'' \emph{IEEE
  Transactions on Information Theory}, vol.~46, no.~2, pp. 325--343, 2000.

\bibitem{frey1997factor}
B.~Frey, F.~Kschischang, H.~Loeliger, and N.~Wiberg, ``{Factor graphs and
  algorithms},'' in \emph{The Annual Allerton Conference on Communication
  Control and Computing}, vol.~35.\hskip 1em plus 0.5em minus 0.4em\relax
  Citeseer, 1997, pp. 666--680.

\bibitem{proakis2001digital}
J.~Proakis and M.~Salehi, \emph{{Digital communications}}.\hskip 1em plus 0.5em
  minus 0.4em\relax McGraw-hill New York, 2001.

\bibitem{needell2009cosamp}
D.~Needell and J.~Tropp, ``{CoSaMP: Iterative signal recovery from incomplete
  and inaccurate samples},'' \emph{Applied and Computational Harmonic
  Analysis}, vol.~26, no.~3, pp. 301--321, 2009.

\bibitem{karabulut2004sparse}
G.~Karabulut and A.~Yongacoglu, ``{Sparse channel estimation using orthogonal
  matching pursuit algorithm},'' in \emph{IEEE 60th Vehicular Technology
  Conference}, 2004, pp. 3880--3884.

\end{thebibliography}
\end{document}